\documentclass[12pt, a4paper]{article}
\usepackage[latin1]{inputenc}
\usepackage{vmargin}
\usepackage{amsfonts}
\usepackage{amssymb}
\usepackage{epstopdf}
\usepackage[dvips]{graphicx}
\usepackage{graphicx}
\usepackage{amsmath}
\usepackage{amsthm}
\usepackage{setspace}
\usepackage{multirow}
\usepackage{hyperref}
\usepackage{cases}
\usepackage[bottom]{footmisc}
\usepackage[small]{caption}

\newtheorem{Proposition}{Proposition}

\begin{document}

\title{Execution and block trade pricing with optimal constant rate of participation\thanks{This research has been conducted with the support of the Research Initiative ``Exécution optimale et statistiques de la liquidité haute fréquence'' under the aegis of the Europlace Institute of Finance. I would like to thank Nicolas Grandchamp des Raux (HSBC France), Jean-Michel Lasry (Université Paris-Dauphine), Guillaume Royer (Ecole Polytechnique) and Christopher Ulph (HSBC) for the conversations we had on the subject. A special thank goes to Charles-Albert Lehalle (Cheuvreux) for having introduced me to the topic of optimal execution many years ago.}}

\author{Olivier Gu\'eant\footnote{Universit\'e Paris-Diderot, UFR de Math\'ematiques, Laboratoire Jacques-Louis Lions. 175, rue du Chevaleret, 75013 Paris, France. \texttt{olivier.gueant@ann.jussieu.fr}}
}

\date{}
\maketitle

\begin{center}
\textbf{Abstract}
\end{center}

When executing their orders, investors are proposed different strategies by brokers and investment banks. Most orders are executed using VWAP algorithms. Other basic execution strategies include POV (also called PVol) -- for percentage of volume --, IS -- implementation shortfall -- or Target Close. In this article dedicated to POV strategies, we develop a liquidation model in which a trader is constrained to liquidate a portfolio with a constant participation rate to the market. Considering the functional forms commonly used by practitioners for market impact functions, we obtain a closed-form expression for the optimal participation rate. Also, we develop a microfounded risk-liquidity premium that permits to better assess the costs and risks of execution processes and to give a price to a large block of shares. We also provide a thorough comparison between IS strategies and POV strategies in terms of risk-liquidity premium.

\section*{Introduction}

Stock traders buy and sell large quantities of shares and cannot ignore the significant impact their orders have on the market. In practice, traders face a trade-off between price risk on the one hand and both execution cost and market impact on the other hand. Traders liquidating too fast indeed incur high execution costs but being too slow exposes the trader to possible adverse price fluctuations, effectively leading to liquidation at lower-than-expected prices. For that reason, traders usually split their large orders into smaller ones to be executed progressively. Research on optimal execution -- or optimal liquidation -- mainly focuses on this issue of optimally splitting those large orders.\\

To provide an optimal rhythm for the liquidation process, the most classical framework is the one developed by Almgren and Chriss in their seminal papers \cite{almgren1999value,almgren2001optimal,almgren2003optimal}. This framework has largely been used and enriched either to better fit real market conditions or to enlarge the scope of modeling possibilities. Black-Scholes dynamics for the price has been considered.\footnote{For short periods of time there is no real difference between Bachelier and Black-Scholes models.} Attempts to generalize the model to take account of stochastic volatility and liquidity were made -- \cite{almgren2011optimal}. Discussions on the optimization criterions and their consequences on optimal strategies are also present in the literature (see for instance  \cite{almgren2007adaptive}, \cite{forsyth2009optimal}, \cite{lorenz2010mean} and \cite{tse2011comparison}). The CARA (or Mean-variance) framework is predominant in the literature and it has been studied for instance in \cite{schied2010optimal}, and in \cite{gueantnew} that also considers block trade pricing. Very interesting results in the case of IARA and DARA utility functions are presented in \cite{schied2009risk}. Following the seminal paper by Obizhaeva and Wang \cite{obizhaeva2005optimal}, many authors also tried to model market impact in a different fashion, using transient marjet impact models. Eventually, the literature recently went beyond the question of the optimal rhythm and focused on the tactical layer, that is on the actual way to proceed, using for instance dark pools \cite{kratz2009optimal, kratz2012optimal, laruelle2011optimal} or limit orders \cite{bayraktar2012liquidation, gueant2012general, gueant2012optimal}.\\

Most of the articles in the literature, be they dedicated to the strategic layer (optimal scheduling) or to the tactical layer (liquidation over short slices of time), focus on IS strategies.\footnote{We ignore here the literature on VWAP strategies that is rather orthogonal to the classical literature on optimal execution.} In this article, we consider strategies constrained to have a constant rate of participation to the market. These execution strategies, called POV or PVol strategies, are more common in practice than IS strategies, although they are suboptimal. Strangely, they are not dealt with in the literature and the goal of this paper is to fill in the blank. Instead of choosing a trading curve as in Almgren-Chriss-like models for IS strategies, we optimize over one single parameter: the participation rate. Noticeably, for most functional forms used in practice for the execution cost function, the optimal participation rate can be found in closed form. This is interesting for at least three reasons. First, for trading, an optimal participation rate is easy to communicate on and does not need any complex tool to be used in practice as opposed to the trading curves of most IS strategies. Second, the formula obtained is a function of risk aversion and it can then be inverted to implicit risk aversion from traders' behavior. Third, the closed-form formula obtained for the optimal participation rate permits to write in closed-form a risk-liquidity premium for block trades. In effect, transactions involving large blocks of shares cannot be based on Mark-to-Market (MtM) prices and we provide a microfounded risk-liquidity premium to be added or subtracted to MtM values. Risk-liquidity premia being already known for IS strategies (see \cite{gueantnew}), we provide a comparison between POV-based liquidity premia and IS-based liquidity premia.\\

In Section 1, we present the setup of the model. In Section 2, we compute in closed-form the optimal participation rate of a POV strategy and the associated risk-liquidity premium. We then discuss the results and analyze the influence of the parameters. In Section 3, we provide numerical examples to illustrate our model.

\section{Setup of the model}

Let us fix a probability space $(\Omega, \mathcal{F}, \mathbb{P})$ equipped with a filtration $(\mathcal{F}_t)_{t\in \mathbb{R}_+}$ satisfying
the usual conditions. We assume that all stochastic processes are defined on $(\Omega, \mathcal{F},(\mathcal{F}_t)_{t\in \mathbb{R}_+}, \mathbb{P})$.\\

We consider a trader with a portfolio containing $q_0 > 0$ shares of a given stock\footnote{The case $q_0 < 0$ can be treated using the same tools.} and we assume that he is willing to unwind his portfolio. The velocity at which liquidation is carried out depends on market conditions. Among them, market volume usually has an important role and we introduce a market volume process $(V_t)_{t \in \mathbb{R}_+}$ assumed to be continuous, deterministic,\footnote{The assumption of a deterministic dynamics may seem odd. Practitioners usually consider market volume curves determined statistically to account for the daily seasonality of market volume. A multiplicative factor may then be added depending on the expected market activity, but it is usually deterministic.} and such that $\exists \underline{V} > 0, \overline{V} > 0, \forall t \in \mathbb{R}_+, \underline{V} \le V_t \le \overline{V}$.\\

To model liquidation, we introduce an inventory process $(q_t)_{t \in \mathbb{R}_+}$ by:
$$\forall t \in \mathbb{R}_+, \quad q_t = q_0 - \int_0^t v_s ds,$$
where the strategy $(v_s)_{s\in \mathbb{R}_+}$ belongs to one of the following admissible sets:

\begin{itemize}
  \item Either:
  $$\mathcal{A}_{IS,T} = \Bigg\lbrace (v_t)_{t\in \mathbb{R}_+}, \mathrm{\;progressively\;measurable}, \forall t, v_t \ge 0, \int_0^T v_s ds = q_0 \mathrm{\;a.s.} \Bigg\rbrace,$$
  if one wants to model liquidation using an IS strategy over the time window $[0,T]$. This is the classical Almgren-Chriss framework \cite{almgren1999value,almgren2001optimal,almgren2003optimal} (see also \cite{gueantnew, schied2010optimal}).

  \item Or: $$\mathcal{A}_{POV} = \left\lbrace (v_t)_{t\in \mathbb{R}_+}, \exists \rho> 0, \forall t\ge 0, v_t = \rho V_t 1_{\{\int_0^t \rho V_s ds \le q_0\}}\right\rbrace,$$
  if one wants to model a POV strategy in which the volume traded by the trader is assumed to be proportional to the market volume process: the participation rate being $\rho$.
\end{itemize}

In both cases, the problem faced by the trader is a trade-off between price risk, encouraging to trade fast, and execution cost / market impact, encouraging to unwind the position slowly.\\

We consider that trades impact market prices in two distinct ways. Firstly, there is a permanent market impact (assumed to be linear\footnote{See \cite{gatheral2010no}.}) that imposes a drift to the price process $(S_t)_{t\in \mathbb{R}_+}$:

$$dS_t = \sigma dW_t - k v_t dt , \qquad \sigma >0, k \ge 0.$$

Secondly, the price obtained by the trader at time $t$ is not $S_t$ because of what is usually called instantaneous market impact (or execution costs). To model this, we introduce a function $L \in C(\mathbb{R}_+, \mathbb{R})$ verifying the following hypotheses:\footnote{We want to cover the cases $L(\rho) = \eta \rho^{1+\phi}$ for $\eta >0$ and $\phi >0$.}

\begin{itemize}
  \item $L(0) = 0$,
  \item $L$ is increasing,
  \item $L$ is strictly convex,
  \item $\lim_{\rho \to +\infty} \frac{L(\rho)}{\rho} = +\infty$.
\end{itemize}

This allows to define the cash process $(X_t)_{t \in \mathbb{R}_+}$ as:

$$X_t = \int_0^t \left(v_s S_s - V_s L\left(\frac{v_s}{V_s}\right) - \psi v_s \right) ds,$$
where the execution cost is divided into two parts: a linear part that represents a fixed cost ($\psi \ge 0$) per share -- linked to the bid-ask spread for instance --, and a strictly convex part modeled by $L$.\\

One of the main goal of this paper is to maximize over $v \in \mathcal{A}_{POV}$ the objective function

$$J(v) = \mathbb{E}\left[-\exp(-\gamma X_T)\right],$$ where $T$ is such that $\int_0^T v_s ds = q_0$ and $\gamma>0$ is the absolute risk aversion parameter of the trader.\\

\section{Solution of the problem and block trade pricing}

\subsection{Optimal participation rate}

To solve our optimization problem, a first step consists in computing the value of the cash process during the liquidation process:

\begin{Proposition}
\label{dist}
Let us consider $\rho >0$ and $T$ implicitly defined by $\int_0^T \rho V_s ds = q_0$.\\
Let us then consider $v \in \mathcal{A}_{POV}$ defined by $\forall t \in \mathbb{R}_+, v_t = \rho V_t 1_{t \le T}$.\\
We have:
$$X_T = q_0 S_0 - \psi q_0 - \frac{k}{2} q_0^2 - \frac{L(\rho)}{\rho} q_0 + \sigma \rho \int_0^T \int_t^T V_s ds   dW_t$$
In particular, $X_T$ is normally distributed with mean $$q_0 S_0 - \psi q_0 - \frac{k}{2} q_0^2 - \frac{L(\rho)}{\rho} q_0$$ and variance
$$\sigma^2 \rho^2 \int_0^T \left(\int_t^T V_s ds\right)^2   dt.$$
\end{Proposition}

\emph{Proof:}\\

By definition,
\begin{eqnarray*}
  X_T &=& \int_0^T v_s S_s ds - \int_0^T V_s L\left(\frac{v_s}{V_s}\right)ds  - \psi \int_0^T v_s ds  \\
   &=& q_0 S_0 - q_TS_T - k \int_0^T v_s q_s ds + \int_0^T \sigma q_s dW_s - L\left(\rho\right) \int_0^T V_s ds - \psi \int_0^T v_s ds  \\
   &=& q_0 S_0 - \frac{k}{2} \left(q_0^2 - q_T^2\right) - L\left(\rho\right) \int_0^T V_s ds + \int_0^T \sigma q_s dW_s - \psi q_0.\\
\end{eqnarray*}
Now,
$$q_t  = q_0 - \rho \int_0^t V_s ds = \rho \int_0^T V_s ds - \rho \int_0^t V_s ds = \rho \int_t^T V_s ds,$$
and therefore $$X_T = q_0 S_0 - \psi q_0 - \frac{k}{2} q_0^2 - \frac{L\left(\rho\right)}{\rho} q_0 + \sigma \rho \int_0^T \int_t^T V_s ds   dW_t.$$

Since $(q_t)_t$ is deterministic, we obtain that $X_T$ is normally distributed with mean $$q_0 S_0 - \psi q_0 - \frac{k}{2} q_0^2 - \frac{L(\rho)}{\rho} q_0$$ and variance
$$\sigma^2 \rho^2 \int_0^T \left(\int_t^T V_s ds\right)^2   dt.$$\qed\\

This proposition permits to write the objective function $J$ in closed-form:

\begin{Proposition}
\label{objective}
Let us consider $\rho >0$ and $T$ implicitly defined by $\int_0^T \rho V_s ds = q_0$.\\
Let us then  consider $v \in \mathcal{A}_{POV}$ defined by $\forall t \in \mathbb{R}_+, v_t = \rho V_t 1_{t \le T}$.\\
$$J(v) = -\exp\left(-\gamma\left(q_0 S_0 - \psi q_0 - \frac{k}{2} q_0^2 - \frac{L(\rho)}{\rho} q_0 - \frac \gamma 2 \sigma^2 \rho^2 \int_0^T \left( \int_t^T V_s ds\right)^2 dt\right)\right).$$
\end{Proposition}

\emph{Proof:}\\

Using Proposition 1, we know that $$\mathbb{E}\left[-\exp(-\gamma X_T)\right] = -\exp\left(-\gamma \mathbb{E}[X_T] - \frac {\gamma^2} 2 \mathbb{V}[X_T] \right) $$$$=-\exp\left(-\gamma\left(q_0 S_0 - \psi q_0 - \frac{k}{2} q_0^2 - \frac{L(\rho)}{\rho} q_0 - \frac \gamma 2 \sigma^2 \rho^2 \int_0^T \left( \int_t^T V_s ds\right)^2 dt\right)\right).$$\qed\\

A consequence of this proposition is that the problem boils down to minimizing:

$$\begin{array}{ccccl}
\mathcal{J} &:& \mathcal{R}_+^* &\to& \mathbb R\\
 & & \rho & \mapsto & \frac{L(\rho)}{\rho} q_0 + \frac \gamma 2 \sigma^2 \rho^2 \int_0^T \left( \int_t^T V_s ds\right)^2 dt, \mathrm{\;where\;} T \mathrm{\;satisfies\;} \rho \int_0^T V_s ds = q_0.\\
\end{array}$$

\begin{Proposition}
\label{existence}
There exists $\rho^* >0$ such that $\mathcal{J}$ has a global minimum in $\rho^*$.
\end{Proposition}

\emph{Proof:}\\

To prove this results, we just need to prove that $\lim_{\rho \to 0} \mathcal{J}(\rho) = \lim_{\rho \to +\infty} \mathcal{J}(\rho) = +\infty$.\\

We know that $L$ has superlinear growth. Hence $\lim_{\rho \to +\infty} \frac{L(\rho)}{\rho} q_0  = +\infty$ and therefore $\lim_{\rho \to +\infty} \mathcal{J}(\rho) = + \infty$.\\

As far as the limit in $0$ is concerned, let us notice that:

$$\int_0^T \left( \int_t^T V_s ds\right)^2 dt \ge \underline{V}^2 \frac{T^3}3.$$
Also, $$\rho^2 = \frac{q_0^2}{\left( \int_0^T V_s ds\right)^2} \ge \frac{q_0^2}{\overline{V}^2 T^2}.$$

Hence, $$\rho^2 \int_0^T \left( \int_t^T V_s ds\right)^2 dt \ge \frac{\underline{V}^2}{3\overline{V}^2} T \to_{\rho \to 0} +\infty.$$

This gives $\lim_{\rho \to 0} \mathcal{J}(\rho) = +\infty$.\qed\\

Proposition \ref{existence} only states that there exists an optimal rate of participation. Interestingly, when one considers the execution cost functions used in practice and approximates the volume curve by a flat volume curve, there is a unique constant participation rate that can be obtained in closed-form:

\begin{Proposition}
\label{closed}
Let us consider the special case where:
\begin{itemize}
  \item the execution function $L$ is given by $L(\rho) = \eta \rho^{1+\phi}$,
  \item the market volume is assumed to be constant: $V_t = V$.
\end{itemize}
Then, there is a unique participation rate minimizing $\mathcal{J}$ given by:
$$\rho^* =  \left(\frac{\gamma \sigma^2}{6\eta \phi}  \frac{q_0^2}{V}\right)^{\frac{1}{1+\phi}}.$$
\end{Proposition}

\emph{Proof:}\\

In the special case we consider, we can simplify the expression for $\mathcal{J}$:

$$\mathcal{J}(\rho) = \frac{L(\rho)}{\rho} q_0 + \frac \gamma 2 \sigma^2 \rho^2 V^2 \frac{T^3}3.$$

Now, since $q_0 = \rho V T$, we have:

$$\mathcal{J}(\rho) =\eta \rho^{\phi} q_0 + \frac{\gamma}{6} \sigma^2 \frac{q_0^3}{\rho V}.$$

This function clearly has a minimum at $\rho^*$ given by the first order condition:

$$ \eta \phi {\rho^*}^{\phi-1} = \frac{\gamma}{6} \sigma^2 \frac{q_0^2}{{\rho^*}^2 V},$$
\emph{i.e.}:
$$\rho^* =  \left(\frac{\gamma \sigma^2}{6\eta \phi}  \frac{q_0^2}{V}\right)^{\frac{1}{1+\phi}}.$$
\qed\\

The expression we obtained in Proposition \ref{closed} for the optimal participation rate allows to carry out comparative statics. However, before going into comparative statics, several general remarks deserve to be made. First, the formula for $\rho^*$ is not bounded by $1$. This is natural because we considered a constant participation rate with respect to the market volume that does not take into account our own volume. Second, most of the literature does not say anything about the value of the risk aversion parameter $\gamma$. We see our closed-form formula as a way to implicit $\gamma$ from traders' behavior. Then the formula is useful to be coherent across stocks. Third, the optimal participation rate does not depend on $\psi$, nor on $k$. Both the linear part of the execution cost and the (linear) permanent market impact have indeed to be paid independently of the participation rate.\\

Now, concerning the other parameters, we have the following results:

\begin{itemize}
  \item When the risk aversion parameter $\gamma$ increases, the trader has an incentive to execute faster to reduce price risk.
  \item The same reasoning applies to $\sigma$. If volatility increases, the trader wants to trade faster.
  \item Since a trader with a larger inventory is exposed to more price risk, the optimal participation rate has to be an increasing function of $q_0$.
  \item $V$ measures the overall liquidity of the stock. The instantaneous volume executed by the trader is $\rho^*V =  \left(\frac{\gamma \sigma^2}{6\eta \phi} q_0^2\right)^{\frac{1}{1+\phi}} V^{\frac{\phi}{1+\phi}}$ and this expression is increasing with $V$. It means that the more liquid the stock, the faster we liquidate the portfolio.
  \item $\eta$ is a scale parameter for the execution costs paid by the trader. If $\eta$ increases, the trader liquidates more slowly.
  \item $\phi$ measures the convexity of the execution cost function. As long as we are in the relevant case $\rho^* \le 1$, the above expression for the optimal participation rate is a decreasing function of $\phi$. It means that the more convex $L$ is, the slower the liquidation process. This is in line with the intuition.
\end{itemize}

\subsection{Block trade pricing and risk-liquidity premium}

In addition to the closed form expression for the optimal participation rate, an important question is the total cost of liquidation when one uses the optimal participation rate. The framework we develop permits to give a price to a block trade of $q_0 > 0$ shares and hence to give a price to liquidity. This is done using the notion of certainty equivalent, or equivalently using indifference pricing -- since we are in a CARA framework. We implicitly define the price $P(q_0)$ of a block trade with $q_0 > 0$ shares through the certainty equivalent of $X_T$:

$$\sup_{v \in \mathcal{A}_{POV}}\mathbb{E}\left[-\exp(-\gamma X_T)\right] = -\exp(-\gamma P(q_0)).$$

This gives:

$$ P(q_0)= q_0 S_0 - \psi q_0 - \frac{k}{2} q_0^2 - \inf_{\rho > 0} \mathcal{J}(\rho).$$

The risk-liquidity premium, is then:
$$ \ell_{POV}(q_0) = q_0 S_0 - P(q_0) = \psi q_0 + \frac{k}{2} q_0^2 + \inf_{\rho > 0} \mathcal{J}(\rho).$$

Under the hypotheses of Proposition \ref{closed}, we obtain the price of a block trade in closed-form.

\begin{Proposition}
\label{closedprice}
Let us consider the special case where:
\begin{itemize}
  \item the execution function $L$ is given by $L(\rho) = \eta \rho^{1+\phi}$,
  \item the market volume is assumed to be constant: $V_t = V$.
\end{itemize}
Then:$$P(q_0) = q_0 S_0 - \psi q_0 - k \frac{q_0^2}{2} - (1+\phi)\eta^{\frac{1}{1+\phi}} \left(\frac{\gamma \sigma^2}{6 \phi V}\right)^{\frac{\phi}{1+\phi}} q_0^{\frac{1+3\phi}{1+\phi}}.$$
\end{Proposition}

\emph{Proof:}\\

$\inf_{\rho > 0} \mathcal{J}(\rho) = \mathcal{J}(\rho^*) = \eta {\rho^*}^{\phi} q_0 + \frac{\gamma}{6} \sigma^2 \frac{q_0^3}{\rho^* V}.$ If we plug the expression for $\rho^*$ into this equation we get:

 $$\eta {\rho^*}^{\phi} q_0 = \eta^{\frac{1}{1+\phi}} \left(\frac{\gamma \sigma^2}{6 \phi V}\right)^{\frac{\phi}{1+\phi}} q_0^{\frac{1+3\phi}{1+\phi}}$$ and
 $$ \frac{\gamma}{6} \sigma^2 \frac{q_0^3}{\rho^* V} = \phi \eta^{\frac{1}{1+\phi}} \left(\frac{\gamma \sigma^2}{6 \phi V}\right)^{\frac{\phi}{1+\phi}} q_0^{\frac{1+3\phi}{1+\phi}}$$

  Hence, $$P(q_0) = q_0 S_0 - \psi q_0 - k \frac{q_0^2}{2} - (1+\phi)\eta^{\frac{1}{1+\phi}} \left(\frac{\gamma \sigma^2}{6 \phi V}\right)^{\frac{\phi}{1+\phi}} q_0^{\frac{1+3\phi}{1+\phi}}.$$\qed\\

This proposition permits to write the risk-liquidity premium as:

$$\ell_{POV}(q_0) = k \frac{q_0^2}{2} + \psi q_0 + \eta^{\frac{1}{1+\phi}} \left(\frac{\gamma \sigma^2}{6 \phi V}\right)^{\frac{\phi}{1+\phi}} q_0^{\frac{1+3\phi}{1+\phi}} + \phi \eta^{\frac{1}{1+\phi}} \left(\frac{\gamma \sigma^2}{6 \phi V}\right)^{\frac{\phi}{1+\phi}} q_0^{\frac{1+3\phi}{1+\phi}}$$

The first term corresponds to unavoidable costs, independent of any optimization, linked to permanent market impact. The second and third term corresponds to the execution costs paid when using the optimal constant participation rate $ \rho^*$. The fourth term corresponds to market risk and it is an implicit cost that should be priced when large blocks of shares are traded. This risk-liquidity premium depends on the parameters in the following way:

\begin{itemize}
  \item The higher the risk aversion $\gamma$, the higher the risk-liquidity premium. This is not surprising and almost a direct consequence of the definition of the risk-liquidity premium. The more risk adverse a trader is, the higher risk-liquidity premium he should quote to compensate for the risk.
  \item Similarly, the more volatile the market, the higher the risk-liquidity premium. In a highly volatile market, the trader quotes a high risk-liquidity premium to compensate for price risk.
  \item Due to convexity and superlinearity in liquidation cost, the last two terms exhibit a convex (increasing) and superlinear behavior with respect to $q_0$. This is also the case of the first term linked to permanent market impact.
  \item As far as $V$ is concerned, the more liquid a market, the lower the risk-liquidity premium.
  \item The higher the execution costs (\emph{i.e.} the higher $\eta$), the higher the risk-liquidity premium.
  \item With respect to the degree of convexity $\phi$ of the execution cost function $L$, the risk-liquidity premium turns out to be decreasing as long as $\rho^* \le 1$.
\end{itemize}

Interestingly, we can compare the risk-liquidity premium obtained when liquidation is constrained to be at constant participation rate (POV strategy), and when there is no constraint (IS strategy). In \cite{gueantnew}, the risk-liquidity premium in the case of a time-unconstrained IS strategy is given by:

$$\ell_{IS}(q_0)=q_0S_0 - \lim_{T \to +\infty} -\frac 1 \gamma \log\left(\inf_{v \in \mathcal{A}_{IS,T}} \mathbb{E}\left[\exp(-\gamma X_T)\right]\right)$$$$ =   \frac k2 q_0^2 + \psi q_0 + \frac{(1+\phi)^2}{1+3\phi} \eta^{\frac{1}{1+\phi}} \left(\frac{\gamma \sigma^2}{2\phi V}\right)^{\frac{\phi}{1+\phi}} q_0^{\frac{1+3\phi}{1+\phi}}.$$

The only difference is obviously in the terms linked to the optimization of the liquidation process. Because the liquidation strategy is unconstrained, the last term in the expression of $\ell_{IS}(q_0)$ is lesser than the last two terms of $\ell_{POV}(q_0)$. Interestingly, we can bound from below the ratio of the liquidity premia:

\begin{Proposition}
$$\frac{\ell_{IS}(q_0)}{\ell_{POV}(q_0)} \ge \frac{1+\phi}{1+3\phi} 3^{\frac{\phi}{1+\phi}} \ge \frac{e \log(3)}{2\sqrt{3}} \ge 0.86$$

\end{Proposition}

\emph{Proof:}\\

$$\frac{\ell_{IS}(q_0)}{\ell_{POV}(q_0)} = \frac{\frac k2 q_0^2 + \psi q_0 + \frac{(1+\phi)^2}{1+3\phi} \eta^{\frac{1}{1+\phi}} \left(\frac{\gamma \sigma^2}{2\phi V}\right)^{\frac{\phi}{1+\phi}} q_0^{\frac{1+3\phi}{1+\phi}}}{k \frac{q_0^2}{2} + \psi q_0 + (1+\phi)\eta^{\frac{1}{1+\phi}} \left(\frac{\gamma \sigma^2}{6 \phi V}\right)^{\frac{\phi}{1+\phi}} q_0^{\frac{1+3\phi}{1+\phi}}}$$
$$\ge \frac{\frac{(1+\phi)^2}{1+3\phi} \eta^{\frac{1}{1+\phi}} \left(\frac{\gamma \sigma^2}{2\phi V}\right)^{\frac{\phi}{1+\phi}} q_0^{\frac{1+3\phi}{1+\phi}}}{(1+\phi)\eta^{\frac{1}{1+\phi}} \left(\frac{\gamma \sigma^2}{6 \phi V}\right)^{\frac{\phi}{1+\phi}} q_0^{\frac{1+3\phi}{1+\phi}}} = \frac{1+\phi}{1+3\phi} 3^{\frac{\phi}{1+\phi}}$$

Let us define $g(\phi) = \frac{1+\phi}{1+3\phi} 3^{\frac{\phi}{1+\phi}}$ for $\phi > 0$. $g$ is a U-shaped function, the minimum being reached at $\phi^*$ implicitly defined by:

$$g'(\phi^*) = 3^{\frac{\phi^*}{1+\phi^*}} \left(-\frac{2}{(1+3\phi^*)^2} + \frac{\log(3)}{(1+3\phi^*)(1+\phi^*)} = 0 \right)$$

This gives $\phi^* = \frac{2-\log(3)}{3\log(3)-2}$. Now, $g(\phi) \ge g(\phi^*) = \frac{e \log(3)}{2\sqrt{3}}$ and this proves the result.\qed\\

The above Proposition means in practice that the gain in going from a POV execution strategy to an IS execution strategy is bounded from above by $14\%$ in terms of risk-liquidity premium\footnote{This bound depends obviously on our assumptions}. Interestingly, $\phi^* \simeq 0.7$ and this is close to the value usually considered by practitioners.\\

\section{Numerical examples}

To exemplify our model, we compute optimal participation rates and risk-liquidity premia for 6 liquidation cases involving Total, Axa and Danone, using a market impact model calibrated on real transaction data.\\ For each stock, we consider two trades representing respectively $10\%$ and $15\%$ of the average daily volume. The input data table is the following:\footnote{We round figures to ease readability.}

\begin{center}
\begin{tabular}{|c|c|c|c|}
\hline
  & Total & Axa & Danone \\
  \hline
  \hline
  \multicolumn{4}{|c|}{\emph{Characteristics of the stock}}\\
  \hline
  \hline
  Price & 40 & 13 & 50 \\
  \hline
  Average Daily Volume (in million shares) & 4 & 7 & 1.7  \\
  \hline
  Annualized volatility & 18\% & 22\% & 18\% \\
  \hline
  \hline
  \multicolumn{4}{|c|}{\emph{Market impact model}}\\
  \hline
  \hline
  $\eta$ & 0.116 & 0.046 & 0.145 \\
  \hline
  $\phi$& 0.63 & 0.63 & 0.63 \\
  \hline
  $\psi$ & 0.002 & 0.0007 & 0.003 \\
  \hline
  $k$ & $5.8\times10^{-7}$ & $1.9\times10^{-7}$ & $2.7\times10^{-6}$ \\
  \hline
  \hline
  \multicolumn{4}{|c|}{\emph{Risk aversion $\gamma = 3\times10^{-6}$}}\\
  \hline
\end{tabular}
\end{center}

The resulting optimal participation rate $\rho^*$ is given in the following table, along with the POV-based premium $\ell_{POV}$ and the $IS$-based premium $\ell_{IS}$. For the POV-based premium, we decompose it into three parts according to the definition of the risk-liquidity premium:

$$\ell_{POV}(q_0) = \underbrace{k \frac{q_0^2}{2}}_{permanent\; market\; impact} + \underbrace{\psi q_0 + \eta^{\frac{1}{1+\phi}} \left(\frac{\gamma \sigma^2}{6 \phi V}\right)^{\frac{\phi}{1+\phi}} q_0^{\frac{1+3\phi}{1+\phi}}}_{instantaneous\;market\;impact}$$$$ + \underbrace{\phi \eta^{\frac{1}{1+\phi}} \left(\frac{\gamma \sigma^2}{6 \phi V}\right)^{\frac{\phi}{1+\phi}} q_0^{\frac{1+3\phi}{1+\phi}}}_{risk}$$
\begin{center}
\begin{tabular}{|c|c|c|c|c|c|c|}
\hline
  &\multicolumn{2}{|c|}{Total} & \multicolumn{2}{c}{Axa} & \multicolumn{2}{|c|}{Danone}\\\hline\hline
  $q_0$ (with respect to ADV) & $10\%$ & $15\%$ & $10\%$ & $15\%$ & $10\%$ & $15\%$ \\\hline
  Optimal Participation Rate & 17.1\% &28.1\% & 13.7\% & 22.5\% & 11.6\% & 19.1\%  \\\hline
  \emph{Perm. m.i. component (bps)}&29.0  &43.5 & 51.2 & 76.8 & 45.9 & 68.8  \\\hline
  \emph{Inst. m.i. component (bps)}&10.1 &13.6 & 10.7 & 14.4 & 8.0 &  10.8 \\\hline
  \emph{Risk component (bps)}& 6.0  &8.2 & 6.4 & 8.7 & 4.7 & 6.4  \\\hline
  POV-based premium (bps)&45.1 &65.3 & 68.3 & 99.9 & 58.6 & 86.0  \\\hline
  IS-based premium (bps)&43.0 &62.4 & 66.0 & 96.8 & 57.0 & 83.8  \\\hline
\end{tabular}
\end{center}

We see that the permanent market impact component represents a large part of the POV-based premium. Hence, there is only little difference in terms of costs between a POV strategy and an IS strategy for the stocks we consider and for the level of risk aversion we consider.

\section*{Conclusion}

Instead of optimizing over all liquidation trading curves as for a time-unconstrained IS strategy, this paper deals with the optimal rhythm to liquidate a portfolio using a constant participation rate (POV strategy). We showed that for most functional forms used in practice for the execution cost function $L$, a closed-form expression was available for the optimal participation rate. We then derived the price of a block trade in this framework and discuss the difference between the risk-liquidity premium quoted by a trader who trade at the optimal constant participation rate and the risk-liquidity premium quoted by a trader using the classical trading curve of an Almgren-Chriss like model.\\

\end{document}